# Timescales of Cell Membrane Fusion Mediated by SARS-CoV2 Spike Protein and its Receptor ACE2

Dominic Hayward[1], Purushottam S Dubey[1], Marie-Sousai Appavou[1], Olaf Holderer[1], Henrich Frielinghaus[1], Sylvain Prevost[2], Bela Farago[2], Anna Sokolova[3], Piotr Zolnierczuk[4], Heiner von Buttlar[5], Peter Braun[5], Joachim Jakob Bugert[5], Rosina Ehmann[5], Sebastian Jaksch[1,6],*

[1]Forschungszentrum Jülich GmbH, Jülich Centre for Neutron Science (JCNS) at Heinz Maier-Leibnitz Zentrum (MLZ), Lichtenbergstraße 1, 85748 Garching, Germany
[2]Institut Laue-Langevin, 71 Avenue des Martyrs, CEDEX 9, 38042 Grenoble, France
[3]Australia Nuclear Science & Technology Organisation, New Illawarra Road, Lucas Heights, NSW 2234, Australia
[4]Neutron Scattering Division, Oak Ridge National Laboratory, Oak Ridge, TN 37831, USA
[5]Institute for Microbiology, Bundeswehr, 80937 München, Germany
[6]European Spallation Source (ESS) ERIC, Partikelgatan 2, 224 84 Lund, Sweden

*Corresponding author: sebastian.jaksch@ess.eu

## Abstract

*In this manuscript we describe the investigation of the SARS-CoV2 membrane fusion timescale by means of small-angle neutron scattering (SANS) using hydrogen/deuterium contrast variation. After the successful production of virus-like vesicles and human-host-cell-like vesicles we were able to follow the fusion of the respective vesicles in real-time. This was done using deuterated and protonated phospholipids in the vesicles in a neutron-contrast matched solvent. The vesicles were identical apart from either the presence or absence of the SARS-CoV2 spike protein. The human-host-cell-like vesicles were carrying an ACE2 receptor protein in all cases. In case of the absence of the spike protein a fusion over several hours was observed in agreement with literature, with a time constant of 4.5 h. In comparison, there was not time-evolution, but immediate fusion of the vesicles when the spike protein was present. Those two figures, fusion over several hours and fusion below 10 s corresponding to the absence or presence of the spike protein allow an upper-limit estimate for the fusion times of virus-like vesicles with the SARS-CoV2 spike protein of 10 s. This very fast fusion, when compared to the case without spike protein it is a factor of 2500, can also help to explain why infection with SARS-CoV2 can be so effective and fast. Studying spike protein variants using our method may explain differences in transmissibility between SARS-CoV2 strains. In addition, the model developed here can potentially be applied to any enveloped virus.*

## Introduction

The corona virus SARS-CoV-2 and the resulting COVID-19 disease have caused serious harm worldwide, not only in terms of public health but also on a socio-economic level. Despite the continuing vaccination campaign, the appearance and rapid spread of ever more transmissible new variants show that, without an effective treatment and management concept, the virus will remain a significant threat. It is therefore important to investigate and understand the mechanisms that make this virus so effectively transmissible between hosts. The first stage of a viral infection begins with the entry into host cells, a process mediated by interaction of a viral protein with a receptor and subsequent membrane fusion. While there are many approaches to studying viral membrane fusion, they all have individual advantages and disadvantages. Recombinant viruses with reporters like fluorescent proteins allow following infection processes in cell culture [1]. They offer the possibility of tracking the location of the virus and how infection spreads in subsequent infection cycles. However, they cannot provide immediate readout of membrane fusion time scale as the reporter proteins first have to be expressed in the infected cells and yield a time-delayed information where fusion took place. Virus-free cell-cell fusion assays based on cells expressing receptors and viral binding proteins respectively allow for more flexibility but still rely on some kind of fluorescent reporter or chemical or enzymatic substrate reaction [2, 3]. Here we highlight the benefit of combining very different fields of the life sciences bringing together biology and physics to yield new perspectives on how to study membrane fusion. Neutron scattering can be used to detect membrane fusion in near real time. Here we exploited the isotope sensitivity of neutrons, where we can label different constituents using either hydrogen or deuterium, and not change the chemistry of the system by labelling and the inherent property of scattering experiments not to follow the fusion process, such as in an optical microscope, but to measure an ergodic sample. We therefore performed small-angle Neutron Scattering (SANS) experiments with reconstituted viral membrane vesicles and reconstituted human cell vesicles to observe the fusion kinetics of virus-like vesicles (VLVs) with host cell analogue vesicles (HCAVs). The VLVs are derived from the plasma membrane of HEK-293 cell cultures, either with or without corona spike protein (here denominated HEK-S or HEK-N, respectively). The HCAVs on the other hand are created using the plasma membrane of Vero-E6 cell cultures, always expressing the ACE2 protein as a receptor for the spike proteins[4, 5]. Using deuteration of the vesicles we were able to follow the fusion of the vesicles as a function of scattering contrast during time resolved SANS experiments.

The general idea of exchange kinetics between 2 differently labelled entities was already developed by Lund et al.[6]. It describes the decay of initially fully deutrated and protonated entities in a solvent of average scattering length density (SLD). By the time the contrast of both entities decays and the final ideal coherent scattering intensity is zero. From the time evolution of the scattering intensity details about the exchange process can be deducted.

Elucidating the timescale of this initial fusion step both fundamentally increases our ability to understand SARS-CoV-2 infection biology and helps to assess the effectiveness of fusion inhibiting antiviral compounds. It also supports investigations by means of molecular dynamics (MD) simulation [7] by allowing an estimate of the real-world timescale.

To our knowledge, this is the first direct investigation of this kind of the timescale of the membrane fusion mechanism between SARS-CoV2 derived vesicles and human analog vesicles. There should be no fundamental difficulty in transferring this method also to other virus families in the future. Our in-vitro membrane fusion model is a very close analog to the actual conditions occurring during an infection.

We were able to show that the presence of the SARS-CoV2 spike protein increases the fusion speed between HCAVs and VLVs by more than three orders of magnitude when compared to the condition without the SARS-CoV2 spike protein.

## Results

The results described here follow a step wise approach. In order to investigate the fusion between vesicles reconstituted from cell plasma membranes first we had to show that we actually reconstituted the appropriate vesicles, proven by dynamic light scattering (DLS) and cryogenic transmission electron microscopy (Cryo-TEM), and that they contain the proteins which were expressed in the cell cultures. After this reconstitution with the different vesicles (VLVs and HCAVs) we discuss the actual time resolved SANS and neutron spin echo (NSE) experiments.

All experiments were performed in pH adjusted PBS buffer, where the phospholipids were never dried out during the preparation. The sample preparation is described in the Methods section. During the experiments low concentrations (3-5 mg/mL) are used in order to avoid diffusion limitation of the actual fusion process, as well as due to the fact that the lysis process of the plasma membrane could only render small amounts of sample for each run, and needed extensive preparatory work for each individual batch. Also, aliquots were sampled only from a few selected samples, since drying the material to ascertain the concentration resulted in the samples no longer being viable for reconstitution. Until now, studies focused on single peptides instead of the full protein[8]. We however wanted to stay as close to the natural system as feasible and therefore kept the samples in buffer during the complete preparation process. Representative images are shown in Figure 1.

### DLS and TEM

The DLS measurements were performed at room temperature in a Anton-Paar Litesizer 500. In general vesicle sizes of approximately 100 nm from the extracted plasma membrane could be reconstituted, with a reasonably high error of +/- 50 nm. The detailed preparation is published as a separated sample extraction protocol[9]. That fits very well with the goal to arrive at a comparable size to SARS-CoV2 particles, but is of course a lot smaller than a proper human cell in the respiratory tract.

In accordance with the DLS measurements the Cryo-TEM images show vesicles, with a wall thickness on the order of 5-6 nm and a radius of 100 nm. Most of the vesicles are unilamellar, as well as "single vesicles" (i.e. not containaing another vesicle inside).

It should be noted that the pH plays a major role in the size distribution and stability of the vesicles. While at neutral conditions (pH 7.4) the vesicles seem to be stabilized against each other, at lower pH 6.2 the vesicles can come into contact. The iso-electric point for the vesicles used in this study is at approximately pH 5. This is also confirmed by zeta-potential measurements. Overview data is given in Figure 2.

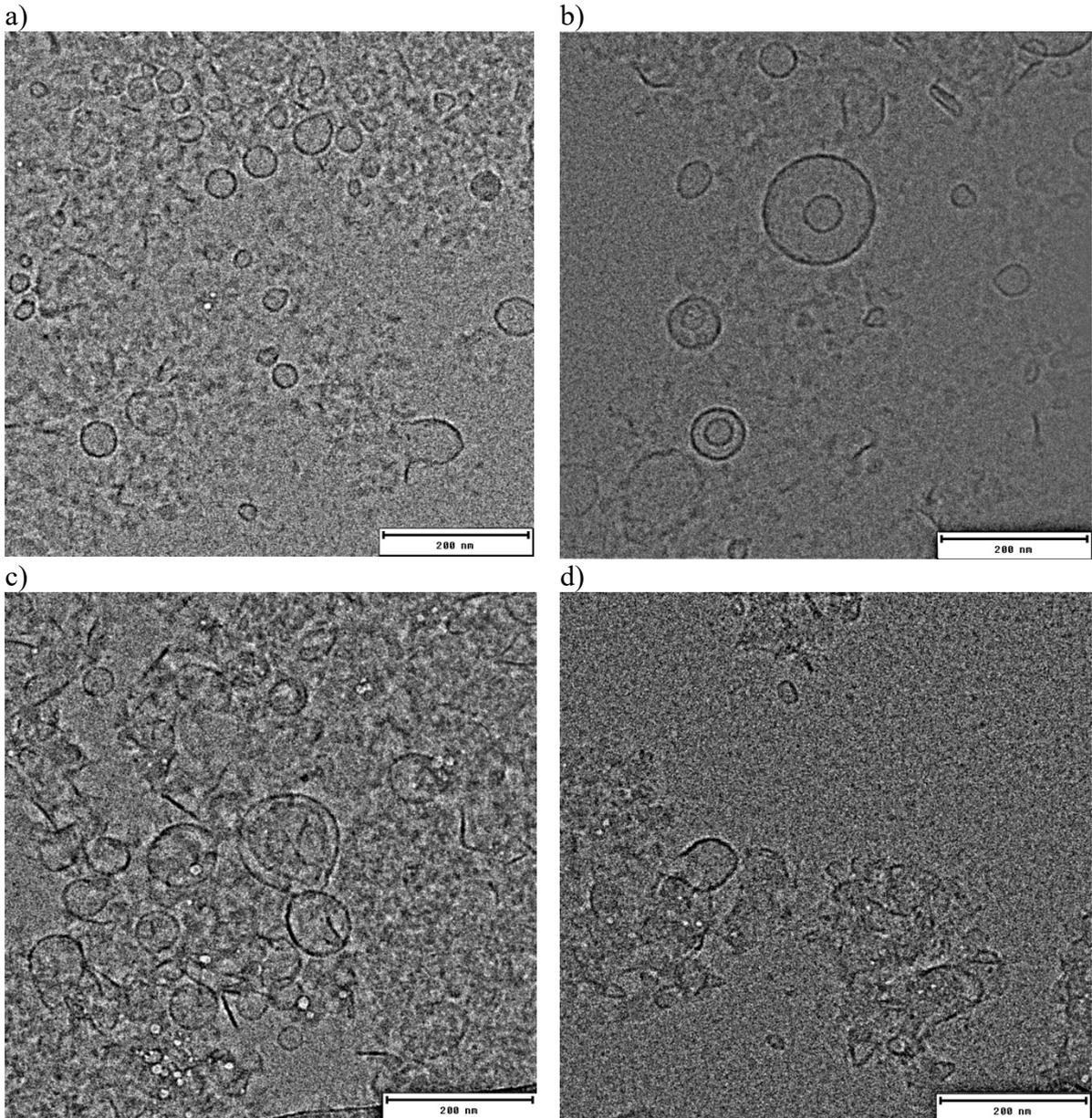

*Figure 1. Representative cryo-TEM images of vesicles after preparation:*

- a) HEK-N @ pH 7.4
- b) Vero @ pH 7.4
- c) HEK-S @ pH 6.15
- d) HEK-N @ pH 5.65

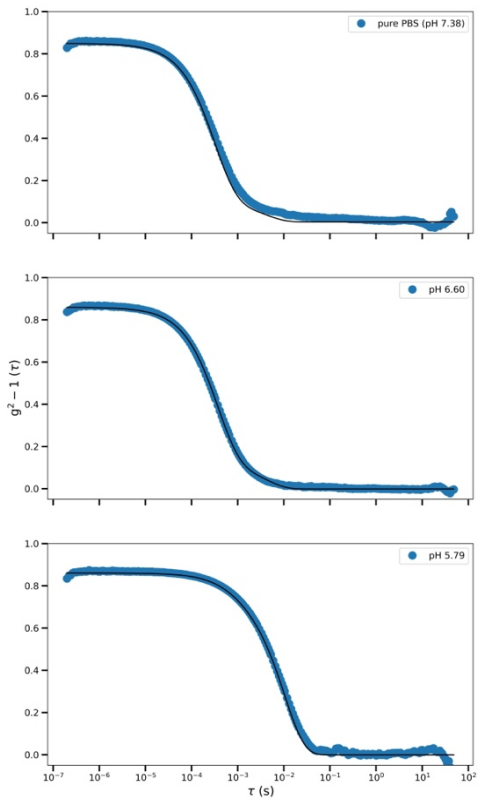 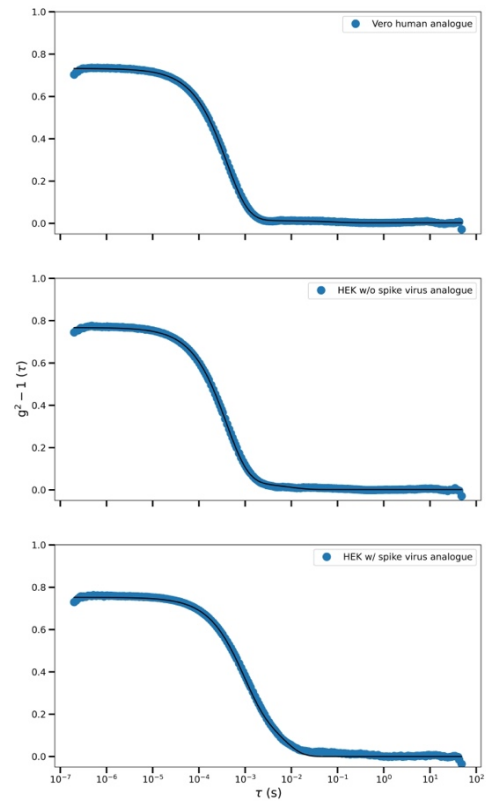

Figure 2. Representative DLS data of vesicles prior to measurements

a) Vero vesicles at different pH

b) (top to bottom, all at pH 6.2): Vero HACV, HEK-N VLV, HEK-S VLV

### Small-Angle Neutron Scattering

We performed experiments both at the Institut Laue-Langevin (ILL), Grenoble, France with the D11 SANS instrument and at ANSTO, Lucas Heights, Australia with the BILBY SANS instrument. We used samples which were partially deuterated, either by growing them in a deuterated medium (ILL), or by mixing with fully deuterated DPPC (ANSTO).

In both cases we deuterated the VLVs derived from the HEK cell line. Since in both cases no full deuteration could be reached, either due to toxicity of the deuterated medium or the need to retain actual sample material in the vesicles, the match point was determined experimentally. The resulting match points were 20 vol% D2O (ILL) and 30 vol% D2O (ANSTO). Both numbers have an error of at least 5% since the number of different concentrations used for SANS was limited by the sample availability and the fact that the match point at 80% $H_2O$ leads to a high background.

All experiments were performed at 37°C set temperature, which resulted in approximately 36.5 °C temperature of the samples during measurement, which matches physiological conditions well. The ILL data was collected with buffers set to pH 7.4 and the ANSTO data was collected at pH 6.2.

We followed the time evolution of the scattered intensity in the forward scattering and used several methods, which all yielded comparable data. For the ILL data, we used the integrated intensity of the forward scattering over five experimental runs where mixing was initiated by the stopped flow setup (BioLogic SFM-3000). The very low sample concentration and the high hydrogen content of the solutions lead to the high measured background which resulted in unreasonably high errors in form factor fitting. In the case of the long-term evolution ANSTO measurements we used a vesicle form factor modeled with SASView [10] and followed the calculated scattering length density of the vesicle shell. Those long-term evolution data were recorded with exposure times of 5 minutes.

Representative data is shown in Figure 3 and Figure 4. Here we have to note the timescales. The ILL data only reaches up to 1800 s, while the ANSTO data reaches well above 10 h. This is due to the fact that we expected fusion to be very fast during our initial experiment, since both the SARS-CoV2 infection is apparently fast, as well as the multitude of proteins would probably result in a lot faster fusion [11] than the pure phospholipid [12]. The times steps are logarithmically spaced for the short time-scale ILL data.

For the ANSTO data however we included both a short time-scale run with the stopped flow setup, as well as slower runs for overnight measurements to also capture long-time kinetics between the vesicles. The fast run reproduced the ILL measurements perfectly.

Due to a longer exposure time, we were able to fit a vesicle form factor to the long time-scale kinetics. For the short time-scale kinetics we only evaluated the inclination exponent of the forward scattering, since the fit could not find stable minima due to the larger error bars. From the time evolution of the SLD of the vesicle shell we can see that in both the long and short timescales we see no time evolution for the samples with Corona spike protein (HEK-S). For the samples without Corona Spike protein (HEK-N) there is no time evolution for the short time-scale kinetics. However, for the long timescales we see a time evolution for the HEK-N samples over several hours with a time constant of 4.5 h. A full exchange, which we attribute with a complete levelling out of the scattering length density, is reached after approximately 7 h.

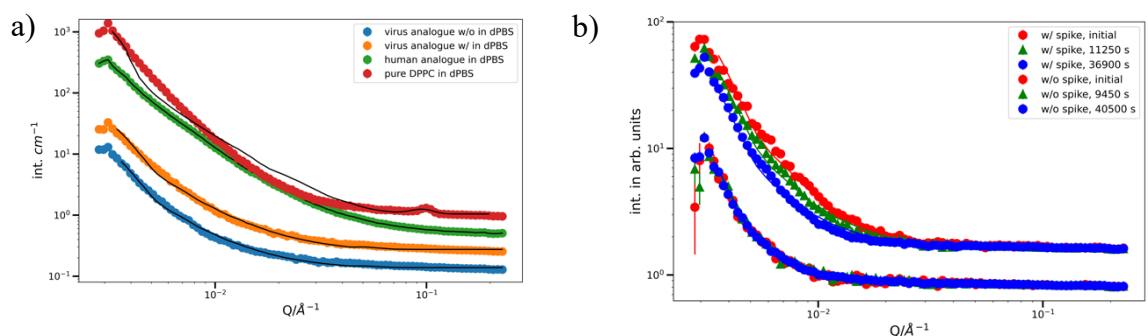

Figure 3. Representative SANS scattering for static measurements and kinetics over long times.

a) SANS curves of static measurements (ANSTO) in deuterated PBS.

b) SANS curves for long period kinetics for fusion with and without SARS-CoV2 spike protein for different times. Curves without spike are shifted by a factor of two against those with spike for better visibility.

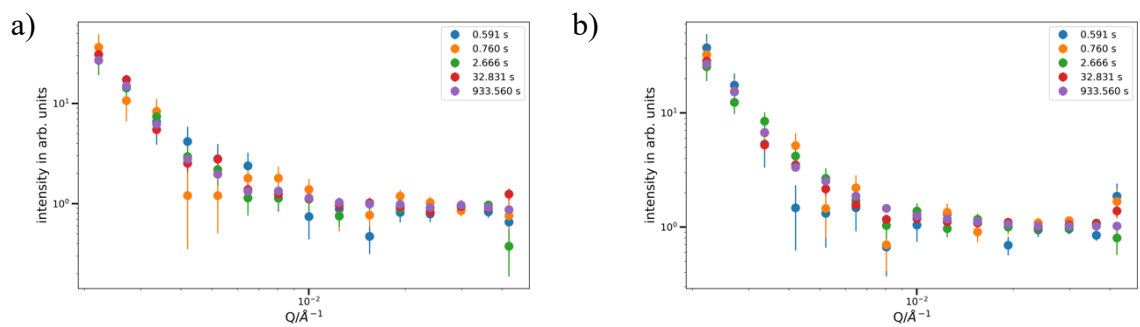

Figure 4. Time evolution for short time-scale kinetics, measured at ILL (a) HEKN (b) HEKS

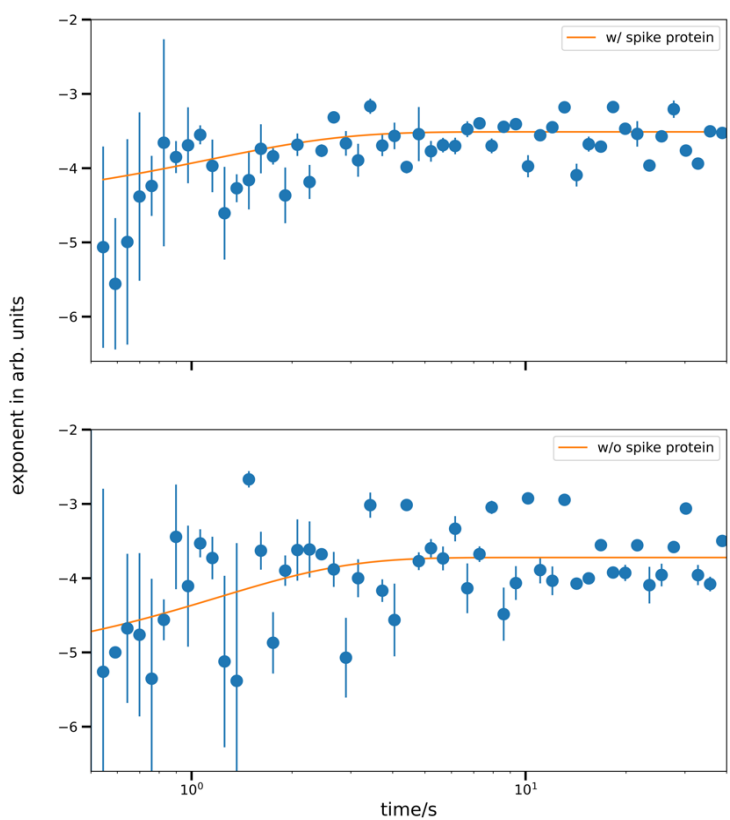

Figure 5. ILL exponent of forward scattering. The huge error bars during fitting as well as the strong scattering of the different values do not allow to fit a time constant for the fusion. The shown fit is a guide for the eye as to what we tried to achieve with this data.

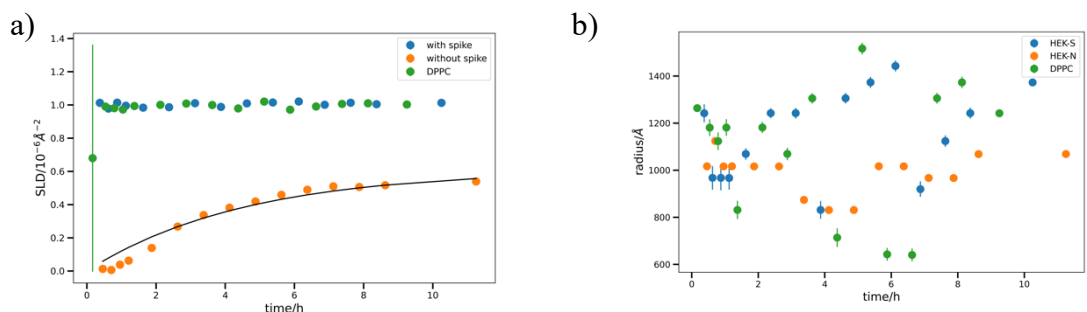

Figure 6. Fitting parameters of vesicle fits to SANS data from long time-scale fusion kinetics.

a) SLD of the shell of vesicles during the fusion process of vesicles with and without spike protein as one fusion partner and vesicles with the ACE2 receptor protein as the other fusion partner. The fitted time constant is 4.53 h.

b) Radius of the corresponding vesicles

Both graphs include error bars. While there is no evolution visible for the vesicle fusion in the presence of the spike protein, there is a distinct time evolution for the fusion without the spike protein, which corresponds well with published data for pure DPPC systems [12]. We attribute this to the fact that with the spike protein the time evolution is too fast to be captured (below 10 s) and can only be observed in the slower case without spike protein. Apart from the protein presence both samples were identical.

## NSE

Neutron spin-echo experiments were performed at the Institut Laue-Langevin (ILL), Grenoble, France with the IN-15 instrument, as well as at the Spallation Neutron Source (SNS) at the Oak Ridge National Laboratory, Oak Ridge/TN, USA with the SNS-NSE instrument.

The diffusion constants as measured by NSE are given in Table 1 and the corresponding measurement data used to obtain the diffusion constant are given in Figure 7. We used SoyPC as a pure phospholipid sample in order to have a baseline for our experiment. Here, the main observation is that the diffusion constant of the VERO membrane is similar to the pure SoyPC vesicle membrane, while HEKS and even more HEKN has a slightly faster decay (i.e. larger diffusion constant). It has to be noted that the diffusion constans of all HEK membranes are similar to the sample with the mixture of HEKS and VERO within the errorbars. Thus, the mixture shows approximately the relaxation time as the HEK membranes alone.

Table 1. Effective diffusion constant from simultaneous fits to the intermediate scattering function.

| Sample | $D_{eff}$ (Å$^2$/ns) |
|---|---|
| SoyPC | 0.86 +/- 0.01 |
| Vero | 1.05 +/- 0.03 |
| HEKS | 2.66 +/- 0.17 |
| HEKS+Vero | 2.72 +/- 0.13 |
| HEKN | 3.02 +/- 0.20 |

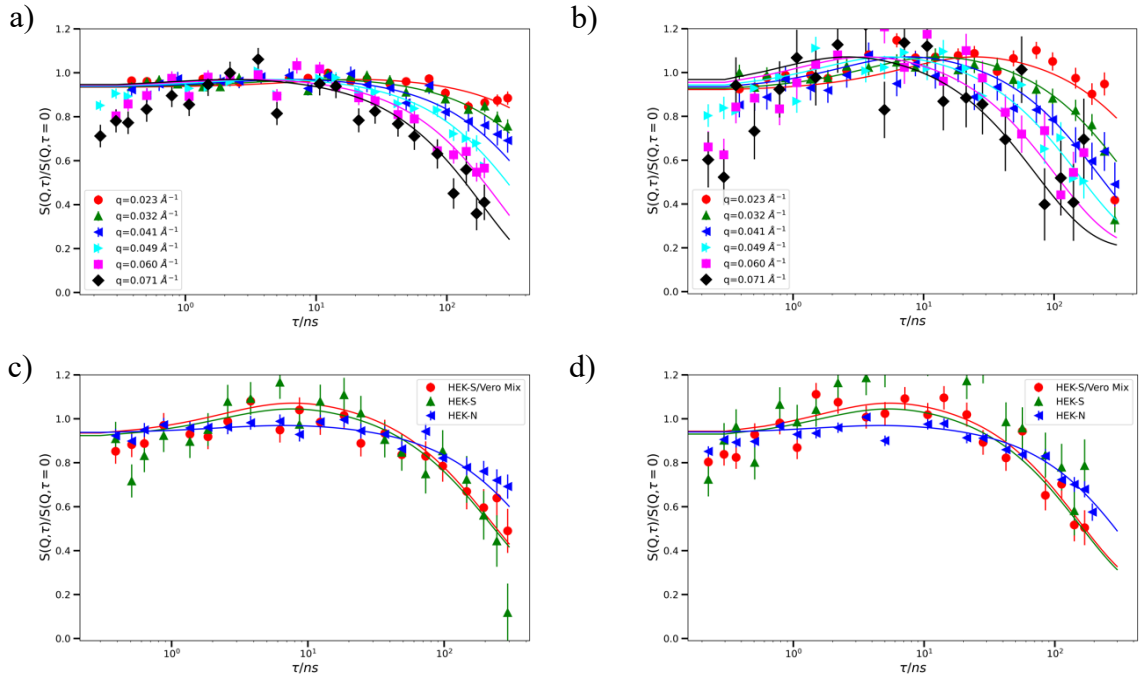

*Figure 7. Intermediate scattering function S(Q, τ) of VERO with simultaneous fits of a) all Q-values , and b) the same for a mixture of HEKS and VERO and comparison of VERO, HEKS and the mixture of both, for c) Q=0.041 Å$^{-1}$ and d) Q=0.049 Å$^{-1}$.*

## Discussion

Here we analyze the time difference in kinetics of the fusion between human analog vesicles HCAVs and virus like vesicles VLVs in the presence or absence of the SARS-CoV2 spike protein. For this we analyze the time difference in the kinetics of the development of the scattered intensity at lower angles (for the short time-scale kinetics) as well as the development of the scattering length density of the shell of the vesicles (for the long time-scale kinetics).

Not seeing a time evolution for the VLVs with spike protein at a first glance seems in both cases an unfeasible experiment was performed. However, there are three points which invalidate this naïve approach: (1) There is a time evolution for the HEKN samples without the corona spike protein, (2) it is completely unlikely that fusion should be completely inhibited when comparing with literature [12, 13] and (3) a very fast fusion was already suspected [7], but our time resolution is limited to 5-10 s, due to the low achievable concentration (3-5 mg/mL) and therefore low scattered intensity.

Building on those three fundamental assumptions we conclude from our data, that the fusion mediated by the SARS-CoV2 spike protein is well below our time resolution of 10 s and the full fusion without the spike protein is on the order of 25 000 s / 7 h, i.e. at least 2 500 times faster (see Figure 5 and Figure 6).

Concerning the pH we observed that neutral conditions seem to stabilize the vesicles. While detailed scans of the zeta potential were not possible due to low amounts of available sample, these scans show a slight adaption to an acidic environment, which is reasonable if when considering the physiological environment in the upper respiratory tract [14].

Based on those results we can extrapolate to the fusion behavior of SARS-CoV2 virions with human host cells. For that we need to consider the differences and similarities between our model system and the actual conditions during an infection. In terms of temperature and pH we were able to replicate physiological conditions quite closely. Also, the nature of the vesicles is as close as possible to the real biological system, directly derived from cell cultures and using the full plasma membrane. This means the vesicles contain both the receptor/acceptor proteins of focus but also the other plasma membrane proteins as well as the phospholipid composition of the plasma membrane. The major difference we have to consider is the amount of sample available. As already mentioned, we used low concentrations of the vesicle preparations (3-5 mg/mL) to avoid crowding in the sample. Under real life conditions, both the human cells are more abundant in the system, more closely resembling a cell lawn in the respiratory tract, as well as the concentration of virions may be higher than the concentration of VLVs in our samples. This restriction however is by design, since we want to avoid crowding in the sample, or diffusion limit effects. Thus, the time scale we observe here is strictly valid only for the first fusion mechanism of a single vesicle.

A final consideration of the applicability of the model is concerning the interplay between the RNA-core of SARS-CoV2 and its envelope. This is to say, since only empty VLVs are considered here, any capsid protein interaction during the fusion cannot be considered.

This timescale of below 10 s for the initial fusion process with spike protein and 4.5 h without spike protein shows an acceleration factor of at least 2500 between the presence and absence of fusion promoting spike protein. This can be interesting as boundary conditions for MD simulations. [7] The appropriate time scales derived from the biological system may improve the predictive power of the MD simulations.

Studying spike protein variants using our method may also explain differences in transmissibility between SARS-CoV strains. The method will also allow to investigate the activity of fusion inhibiting compounds that could be used for antiviral treatment.

The methods presented here can in principle be applied to any other enveloped virus. The authors aim to further address the question of virus fusion and efficacy of antiviral compounds using this method with relevant viruses in future work.

# Methods

### Cells and cell culture

Vero E6 cells were used to produce acceptor vesicles carrying the ACE2 receptor. For the donor vesicles, HEK 293T cells were used to prepare negative control negative control vesicles and HEK 293T cells expressing SARS-CoV-2 spike protein yielded plasma membranes for the vesicles to study SARS-CoV-2 fusion kinetics. All cells were grown in DMEM high glucose (Gibco) with 10 % fetal bovine serum (Gibco) and 1x non-essential amino acids (Gibco). For deuteration of cells the medium was complemented stepwise with medium with deuterated glucose (Sigma) and $H_2O$ stepwise substituted for $D_2O$. Incubation of cells was performed in T75 or T875 vent cap flasks (Falcon) in incubators with 4.5 % $CO_2$ and 90 % humidity at 37 °C. Cells were harvested by washing the cell layers and treating them with trypsin. Cells were detached by gently shaking the flasks and collecting the detached cells in DMEM with 10 % FBS.

### Sample Preparation

The plasma membranes of the harvested cells were lyzed in a newly developed procedure based on a previous protocol by Suski et al.[15]. It is important to note that only low yields of approximately 3-5 mg per 5-10 g of cell cultures could be achieved. This is first obviously due to the fact that most of the cells' mass is not made up from the plasma membrane. Secondly, we would rather avoid additional organelles from the inside of the cells in order to avoid contamination of the phospholipid/cholesterol/protein mixture making up the vesicles. The cell cultures were first lyzed in an isolation buffer solution and later stored in PBS.

Disruption of the plasma cell membrane was done by ultrasonic shock (Vibra Cell VCX 130 PB Sonicator), and the samples were subsequently purified by selective centrifugation. This was done to avoid any chemical stimuli to the proteins during the lysis procedure.

Presence of the ACE2 receptor protein was confirmed with an ELISA (Human ACE2 ELISA Kit, Thermo Fisher).

It is important to note that we only used ultrasonic treatment to arrive at our unilamellar vesicles by ultrasonic homogenization. As for the lysis procedure we wanted to avoid any harsh chemical interaction, and other mechanical methods, such as extrusion, proved unfeasible. Contrast match conditions were at 20% $D_2O$ and 80% $H_2O$.

### DLS

The DLS and zeta potential measurements were performed using an Anton Paar Litesizer 500. All measurements were performed at room temperature. Data analysis was done by fitting triple exponential functions to the correlation data from the measurement. With the derived diffusion constant, the radii of the vesicles were calculated [16].

For the zeta potential measurements, the instrument directly reported the value of the potential.

### TEM

Samples for TEM were prepared by placing a drop of 4 microliter solution on a lacey S7/2 carbon-coated copper grid. After a few seconds, excess solution was removed by blotting with filter paper.

This sample was cryo-fixed by rapid immersing into liquid ethane at -180 °C in a cryo-plunge (EMGP Leica GmbH) where the temperature was set to 20°C and the relative humidity at 85% prior to vitrification. The specimen was inserted into a cryo-transfer holder (HTTC 910, Gatan, Munich, Germany) and transferred to a JEM 2200 FS EFTEM instrument (JEOL, Tokyo, Japan). Examinations were carried out at temperatures around -180°C. The transmission electron microscope was operated at an acceleration voltage of 200 kV. Zero-loss filtered images were taken under reduced dose conditions (<10 000 e-/nm$^2$). All images were recorded digitally by a bottom-mounted 16 bit CMOS camera system (TemCam-F216, TVIPS, Munich, Germany). To avoid any saturation of the gray values, all the measurements were taken with intensity below 15 000, considering that the maximum value for a 16 bit camera is 2^16. Images have been taken with EMenu 4.0 image acquisition program (TVIPS, Munich, Germany) and processed, particularly using band pass filter and improving brightness and contrast, with the free digital imaging processing system Image J [17].

### NSE

Neutron spin echo spectroscopy provides information on the thermally driven equilibrium dynamics on molecular length scales (set by Q) and time scales in the range up to some 100's of nanoseconds.

Vesicle membranes fluctuate driven by thermal motion, which gives rise to a characteristic decay of the intermediate scattering function. The vesicle membrane elasticity is one of the main parameters governing the height fluctuations of the membrane. Diffusion of membrane proteins in the plane of the membrane can contribute additionally to S(Q,t), depending on the details of the contrast. The data are interpreted by a simple diffusion model, i.e. relaxation rate divided by q$^2$, gives the Stokes-Einstein diffusion constant for a simply diffusing object and gives also in more complex environment a good hint on the diffusion and on changes of the dynamics with changes in composition (e.g. of the membrane).

Fluctuating vesicles can also be analyzed by a model from Lisy[18], describing the intermediate scattering function due to shape fluctuations of a droplet. Due to the many parameters of this model, and given the statistics of the data, it is not suited for comparison of different membranes and trends in the elastic constants for the given system. A simultaneous fit with this model with reasonable guesses for the missing parameters leads to bending rigidities around some 10 k$_B$T for the membranes.

The low concentration of membrane material in the sample in the experiments presented here resulted in a low coherent intensity, requiring to add a contribution accounting for the incoherent solvent dynamics (even for the mainly deuterated water solvent).

As a reference, the dynamics of a simple phospholipid vesicle, SoyPC, has been measured as well and analyzed in the same way. Values in the table are simultaneous fits to all Q-values for a given sample in order to reduce the error.

The fit function for the intermediate scattering function,

$$\frac{S(Q,\tau)}{S(Q,0)} = A\left(A_{coh}\ \exp(-D_{eff}q^2\tau) + (1 - A_{coh})(1 - \exp(-D_{water}q^2\tau))\right)$$

The second term comprises the incoherent contribution from water diffusion, which has to be set to 220 $Å^2$/ns [19] and accounts for an initial increase of the intermediate scattering function. The main contribution with amplitude $A_{coh}$ represents the main long-term decay, the effective diffusion constant $D_{eff}$ is determined by a simultaneous fit of all Q, as stated above.

We performed experiments both at the Institut Laue-Langevin (ILL), Grenoble, France with the IN15 Neutron Spin Echo (NSE) Spectrometer [20], and at the SNS with the SNS-NSE [21]. The normalized intermediate scattering function $S(Q,\tau)/S(Q,\tau=0)$ (or time correlation function in reciprocal space) is measured by encoding and decoding the neutron velocity of each neutron of a polarized beam by a number of spin precessions in a magnetic field before and after the scattering event at the sample. With this method one keeps track of velocity changes experienced during scattering at the sample, resulting in a reduced polarization of the neutron beam, which is directly related to the intermediate scattering function.

Three wavelength settings (8, 10 and 12 Å) were used at the instrument IN15 with a wavelength spread $\delta\lambda/\lambda=15\%$. A Fourier time range up to 335 ns and a Q-range of 0.023 – 0.071 $Å^{-1}$. Background correction with a $D_2O$ measurement and normalization with a Grafoil resolution has been performed.

At the SNS-NSE two wavelength band settings (5-8 Å, with scattering angles 2θ 3.65° and 6.66° and 8-11 Å with 2θ 5.43°) were used. Due to the low coherent scattering intensity and the remaining incoherent intensity of $D_2O$, the data allowed to evaluate only the lower Q-range, which had been binned into 4 slices from 0.041 to 0.07 $Å^{-1}$. $D_2O$ background has been measured under the same conditions. The background was subtracted from the data. A reference measurement with an elastic scatterer ($Al_2O_3$ and Grafoil) has been conducted for normalization.

## SANS

We performed experiments both at the Institut Laue-Langevin (ILL), Grenoble, France with the D11 SANS instrument and at ANSTO, Lucas Heights, Australia with the Bilby SANS instrument.

The settings at Bilby were 6 Å wavelength with a nominal resolution of 10%. Collimation was set to 14.774 m with the detector at 16.001 m and additional detector curtains at 4.202 and 3.198 m. Separations between the curtains on the left side were 0.2 m, on the right side 0.350 m, 0.05 m at the top and 0.150 m at the bottom. The chosen source and sample apertures were 40.0 and 12.5 mm.

At D11 a wavelength of 4.6 Å with a collimation length of 38 m and a sample detector distance of 45.5 m was chosen.

All experiments were performed at 37°C and at pH 7.4 (ILL) and pH 6.2 (ANSTO).

For the fast time-scale kinetics a stopped flow setup by Biologic was used (SFM-3000). The samples were preheated to 37°C and then injected into the preheated sample chamber for mixing. All shots were repeated at least 5 times. After the measurement an average over all injections was done for each sample composition.

For the slow kinetics at ANSTO we injected the sample by hand and started the measurement manually into 1 mm flight path Hellma Quartz cuvettes. The set temperature was also 37 °C. Concerning the sample analysis, we had two different approaches, due to the different deuteration and therefore scattered intensity of the samples at ILL and ANSTO.

At ILL and for the fast time-scale kinetics at ANSTO the scattered intensity after background correction was so small that no useful form factor could be fitted to the data. Since we were interested mostly in the scattering by the form factor at lower Q-values, we integrated the scattered intensity between Q= 0.004 and 0.007 Å$^{-1}$. The resulting intensity value we observed over time. To confirm validity of the procedure, the same analysis has been performed using various Q limit for integration. All obtained values were consistent with each other. An exponential function of the form $I(q) = I_o q^{-\alpha} + const.$ Was fitted and the resulting exponent was plotted over time. While the resulting values were less dependent on a single point of the intensity curve, they still scattered strongly at short times (see Figure 5). From this we concluded that under the given circumstances no time evolution could be determined from the experiment. Further investigation of the TEM-images showed the vesicles were stabilized against fusion at neutral pH values.

The SANS curves measured for the slow kinetics at ANSTO had 5 minutes of exposure time each. Also, here a previous measurement of the isoelectric point allowed for a more sensible setting of the pH value of pH 6.2, since the vesicles had been stabilized at pH 7.4 and the isoelectric point was at pH 5. With these longer exposure times the scattered intensity was sufficient to fit a form factor for a unilamellar vesicle, using the built in form factor with SASView [22]. Since the only parameter of interest was the scattering length density of the vesicle shell as many of the other parameters as possible were determined a-priori (see Table 2). The only free parameters were the overall radius of the vesicle and the scattering length density of its shell.

In order to obtain a time constant, we fitted the generic expression
$$SLD(t) = SLD_0 [1 - \exp(-t/\tau)]$$
Here $\tau$ is the time constant for the mixing of the vesicles. This was applied to the fitted SLD values from a vesicle form factor.

*Table 2. Constants and fitting parameters for the vesicle form factor during the kinetic experiment.*

| PARAMETER | VALUE |
| --- | --- |
| SCALE | 1 |
| BACKGROUND | 0.806 cm$^{-1}$ |
| SLD_SHELL | 1 ×10$^{-6}$ Å$^{-2}$ (starting value) |
| SLD_SOLVENT | 1.803 ×10$^{-6}$ Å$^{-2}$ |
| VOLUME FRACTION | 1 ×10$^{-3}$ |
| RADIUS | 1000 Å (starting value) |
| THICKNESS OF SHELL | 60 Å |

The slow kinetics with 5 min exposure times were fitted manually and checked for consistency, for the fast kinetics every 10$^{th}$ fit out of the 50 was also performed with more free parameters to validate the calculated parameters were still sensible.

# Acknowledgements

Since the research presented here was performed at international facilities in times of pandemic lockdowns, limited availability of materials and reduced operations at large-scale facilities, we are extremely grateful for the outstanding support that has been given by the support staff at both the Institut Laue-Langevin and the ANSTO facility at Lucas Heights. Without this support, the research presented here would not have been possible.
Here we want to name especially Deborah Wakeham at ANSTO.
This research used resources at the Spallation Neutron Source, a DOE Office of Science User Facility operated by the Oak Ridge National Laboratory.

All granted beamtime is gratefully acknowledged.

This work benefited from the use of the SasView application, originally developed under NSF award DMR-0520547. SasView contains code developed with funding from the European Union's Horizon 2020 research and innovation programme under the SINE2020 project, grant agreement No 654000.